\documentclass[journal,compsoc]{IEEEtran}
\usepackage{graphicx}
\usepackage{epstopdf}%
\usepackage[usenames,dvipsnames]{xcolor}\usepackage{graphicx}
\usepackage{url}
\usepackage{multirow} 
\usepackage{array}

\makeatletter
\def\ps@headings{%
\def\@oddhead{\mbox{}\scriptsize\rightmark \hfil \thepage}%
\def\@evenhead{\scriptsize\thepage \hfil \leftmark\mbox{}}%
\def\@oddfoot{}%
\def\@evenfoot{}}
\makeatother
\usepackage[usenames,dvipsnames]{xcolor}

\usepackage{epsfig}
\usepackage{amsmath}%
\usepackage{amsfonts}%
\usepackage{amssymb}%
\usepackage{graphicx}
\usepackage{wrapfig}
\usepackage{subfigure}
\usepackage[all]{xy}
\usepackage{sidecap}
\usepackage{footnote}
\newcommand{\s}[1]{{#1}}

\newcommand{\ignore}[1]{}

\newcommand{\fullppr}[1]{}

\ignore{

\newcommand{\qed}{\nobreak \ifvmode \relax \else
      \ifdim\lastskip<1.5em \hskip-\lastskip
      \hskip1.5em plus0em minus0.5em \fi \nobreak
      \vrule height0.75em width0.5em depth0.25em\fi}
}

\begin{document}

\title{Ethical Considerations when Employing Fake Identities in OSN for Research}

\author{
Yuval Elovici$^{*}$\thanks{$^*$elovici@bgu.ac.il}, Michael Fire$^{*}$\thanks{$^*$michyfi@bgu.ac.il}\thanks{$*$Telekom Innovation Laboratories and Department of Information Systems Engineering, Department of Information Systems Engineering, Ben Gurion University of the Negev}, Amir Herzberg$^{\text{\textdagger}\ddag}$\thanks{$^{\text{\textdagger}}$amir.herzberg@gmail.com}, Haya Shulman$^{\mathsection\ddag}$\thanks{$^{\mathsection}$haya.shulman@gmail.com}\thanks{${\ddag}$Department of Computer Science, Bar Ilan University}\thanks{$^{\mathsection}$Fachbereich Informatik, Technische Universit\"{a}t Darmstadt/EC-SPRIDE}
}

\maketitle

\begin{abstract}
\s{Online Social Networks (OSNs) have rapidly  become a prominent and widely used service}, offering a wealth of personal and sensitive information with significant security and privacy implications. Hence, OSNs are also an important - and popular - subject for research. To perform research based on real-life evidence, \s{however}, researchers may need to access OSN data, such as texts and files uploaded by users and connections among users. This raises significant ethical problems. 
Currently, there are no clear ethical guidelines, and researchers may end up (unintentionally) performing ethically questionable research, sometimes even when more ethical research alternatives exist. For example, several studies have employed ``fake identities'' to collect data from OSNs, \s{but fake identities may be used for attacks and are considered a security issue}. Is it legitimate to use \s{fake identities} for studying OSNs or for collecting OSN data for research?  
We present a taxonomy of the ethical challenges facing \s{researchers} of OSNs and compare different approaches. 
We demonstrate how ethical considerations \s{have been} taken into account in previous studies that used fake identities.
\s{In addition, several possible approaches are offered to reduce or avoid ethical misconducts.}
We hope this work will stimulate the development and use of ethical practices and methods \s{in} the research of online social networks. 
\end{abstract}

\section{Introduction}

{\em Online Social Networks (OSNs)}, often referred to as {\em social media}, have rapidly become an integral part of society\s{. They are} widely used by many and contain a wealth of \s{information, much of it sensitive and personal}. \s{This vast source of pertinent information} allows \s{for} innovative applications of OSNs, and many of these innovations are designed \s{to benefit both} society at large and individuals in particular to identify \s{possible} threats to society or individuals. \s{For example, a study on OSNs} can assist in \s{the} identification of individuals who are inclined to commit suicide or \s{initiate} a terror attack, thus enabling \s{a priori} prevention. On the other hand, there are also many ways in which OSNs may be abused, \s{resulting in inappropriate or harmful actions}. 

The \s{widespread} use of OSNs, along with their potential for beneficial as well as harmful applications, \s{promotes the study} of OSNs and their users.  \s{Much research clearly has} legitimate motivation, such as the development of new products and services, \s{followed by the evaluation} of their acceptance and usage 
\s{Other research focuses on counteracting harmful applications.
In particular, it is important to perform high-quality, experiment-based research to evaluate risks and the effectiveness of different countermeasures (\cite{fire2012social,fire2012strangers,rahman2012efficient}). }

Online social network research often involves measurements \s{within deployed and} operating OSNs, often focusing on the most popular, such as Facebook.\footnote{\url{http://www.facebook.com/}} Such operational research on deployed \s{OSNs seems to be the best - or even only -} method to study important issues which can help \s{in the design and the improvement} of OSN products\s{; in identifying} threats on OSNs and their users, \s{including the} design of defenses\s{; and in}  understanding social and \s{economic phenomena}. 

Much of the \s{current OSN} research focuses on the following two issues: the {\em OSN graph} and the {\em OSN user behavior}. Research on {\em OSN graphs} \s{analyzes the} properties of the OSN relationships graph of connections among OSN users; this can be important for many goals, such as designing a new OSN product. Research on {\em OSN user behavior} \s{focuses on} typical behaviors of OSN users, which can help design and improve products as well as study security and privacy vulnerabilities and defenses. 

\s{The collection} of data on an operational system, involving data related to real users, raises ethical and even legal concerns and dilemmas. 
\cite{eynon2008ethics} identify three basic ethical concepts for research in humans in general, and specifically \s{in relation to the Internet}: 
{\em confidentiality, anonymity,} and {\em informed consent}\s{. These} concepts are at the core of most institutional and professional research governance. 

\cite{CREDS13:HM} consider a related topic \s{through their study of} usability aspects of systems and ethical problems of such research; they, however, do not consider research \s{specifically on OSNs}.
Consider the ethical issues and approaches, \s{for example}, in the field of {\em Internet measurements}, which involves real operation systems, similar to OSN research (\s{especially with respect to} the OSN graph). 
Most \s{individuals} performing Internet research take privacy and security concerns into consideration. Specifically, studies that expose traffic, e.g., to allow further research on it, normally \s{``sanitize''} it by removing any data considered sensitive\s{; 
this is illustrated in the Cooperative Association for Internet Data Analysis (\cite{CAIDATraces}}). Similar sanitization of exposed data may be appropriate for OSN research as well. 

In this paper, we study the ethics of OSN research as well as potential conflicts between ethical compliance  and \s{attaining} social benefits from research.
Note that OSN research tends to be much more active, with potential impact on both users and OSN providers, compared \s{to other forms of research, such as} Internet measurements. 
This \s{prompts} the consideration of an additional ethical concept, which we call {\em avoid disruption and waste}. 

Online social networks raise additional ethical challenges since 
OSNs normally do not allow such information to be freely available due to the privacy concerns of its users and the OSN terms of use. In fact, the ability to share selected information with only a selected set of peers (``friends'') is an important requirement from OSNs.

As a result, \s{much of the research using} OSNs involves different techniques to collect information,  circumventing these OSN limitations. 
This includes ``whitehat'' research for legitimate academic and industrial goals, as well as  ``blackhat/greyhat'' research, whose goal is to extract and exploit sensitive information as well as to actively connect to users, provide (fake) information for different goals, and \s{perform similar malicious activities.} Such ``blackhat/greyhat'' research is conducted by criminals, hacktivists, and even organizations involved in cyber-warfare (terrorists, armies, intelligence and law-enforcement agencies).

Indeed, one of the goals of academic (and some industrial) researchers is \s{precisely} to study vulnerabilities allowing such greyhat/blackhat research and then design improved defenses for OSNs and their users. However, for such research to be realistic, researchers must base it on actual OSN data. 
This raises the type of ethical problems we study in this paper. 
For example, in order to study information leakage by corporate employees in \s{a particular} OSN, researchers may want to employ similar methods that industrial espionage attackers will use. Similarly, studying \s{the} diffusion of information in a social network requires \s{the} monitoring of many OSN members.

An effective and widely used technique to obtain information about \s{an} OSN and its users is \s{to establish OSN connections with many users}, typically by \s{creating} a significant number of OSN accounts under {\em fake, non-existing identities} and using these to connect to other users.
The creation of such accounts \s{has been studied by several} researchers (\cite{boshmaf2011socialbot,elishar2012}).

Fake identities are widely used. It is estimated that more than 8.7\% of the identities in Facebook are fake (\cite{facebook_fake}).  These identities were created for various purposes with both legitimate and malicious intent. Fake identities may be completely fabricated, or \s{they may be} a clone of \s{an actual} identity that exists in the real world (\cite{elishar2012,kontaxis2011detecting}). Recent studies (\cite{boshmaf2011socialbot,elishar2012}) show that users tend to accept friendship requests from people that they do not know both in the virtual and physical world; hence, it is relatively easy to connect fake identities to real identities.

Obviously, the creation of fake identities raises serious ethical, and even legal, concerns.  Other techniques to collect OSN data also raise ethical concerns. For example, suppose a researcher did not use a fake account, but her own account; would it be ethical for her to publish information that other users shared with her? Is this case a personal issue between the specific researcher and the persons who decided to connect to her, or is it an ethical issue that should be a considered when  accepting such a paper for publication? 

In this paper, we investigate the dilemma between two social goals: (1) the above-mentioned privacy and security concerns, and (2) the desire to have reliable experimental research on OSNs for socially beneficial or at least legitimate goals, such as to improve OSN \s{products, services, security, and privacy}. We explore the need for OSN data for experimental research (Section \ref{sec:acquire}), \s{we evaluate} the ethical concerns (Section \ref{sec:ethics}), and \s{we consider} some possible solutions (Section \ref{sc:sol}).

\subsection{Related Work}
In recent years with the increasing number of online sources accessible to researchers, such as web blogs, discussion boards, and online social networks, there \s{have been a} growing number of studies \s{on OSNs} and users' behavior. Several studies \s{have} investigated the ethical aspects of acquiring and analyzing online users' information. \cite{eysenbach2001ethical} and \cite{flicker2004ethical} \s{examined} the ethical issues which may arise when studying online Internet communities. In their research, Flicker et al presented practical guidelines for resolving ethical dilemmas in these types of studies. \cite{thelwall2006web} investigated the ethical dilemmas that arise when researchers use web crawlers to collect information from online sources. \cite{eynon2009new} explored ethical dilemmas, such as \s{the reuse} of data from multiple sources, which may occur when studying online virtual environments. \cite{wilkinson2011researching} \s{discussed the} ethical considerations of extracting personal information from public online sources  for research purposes. According to Wilkinson and Thelwall, researchers do not need to ask permission from the 
text authors when collecting public information\s{; however,} steps for ensuring that the text authors are anonymized need to be taken in academic studies. Subsequently, \cite{rosenberg2010virtual} investigated what data should be perceived as private and what as public\s{; this enables} researchers to determine when their study requires an informed consent. \s{Recently, \cite{lucas2012ethics} demonstrated that ethics does not get enough serious examination in the field of social eco-informatics. 
Lucas recommended that social eco-informatics researchers include more ethical deliberation in their research.}

Indeed, there is a rising awareness \s{regarding} the ethical concerns related to OSN research, but it mainly focuses exposing a user's private data.
In this work, we discuss ethical problems related to the methods which are deployed when conducting research on an OSN, and \s{we also consider} the potential risks and problems which may \s{result, affecting} the OSN and its users as well as \s{possible} third parties.

\subsection{Contributions}
This work presents the first taxonomy of {\em ethical} considerations and problems related to the range of techniques and approaches that are employed in studies of online social networks. This significantly \s{expands upon previously published  works} on ethics in OSN research; for example, we \s{demonstrate that OSN research} can have a detrimental impact not only on OSN users but also on the OSN provider and even those \s{indirectly associated}. 

\subsection{\s{Organization}}

The rest of the paper is structured as follows: Section \ref{sec:acquire} describes how information stored in OSNs can be harvested for research purposes. Section  \ref{sec:ethics} follows with a discussion of the ethical and legal perspectives related to research on OSNs. Section \ref{sc:sol} discusses several possible approaches to \s{warrant the use of fake identities}  while complying with ethical considerations.
Section~\ref{sc:prev} illustrates how ethical considerations were taken in account in previous research that involved fake identities. \s{The paper provides concluding comments} in Section~\ref{sc:conclusions}.     
\section{Acquiring \s{Online Social Network} Information}
\label{sec:acquire}
\begin{figure*}[th!]
	\begin{center}
	\includegraphics[width=0.80\textwidth]{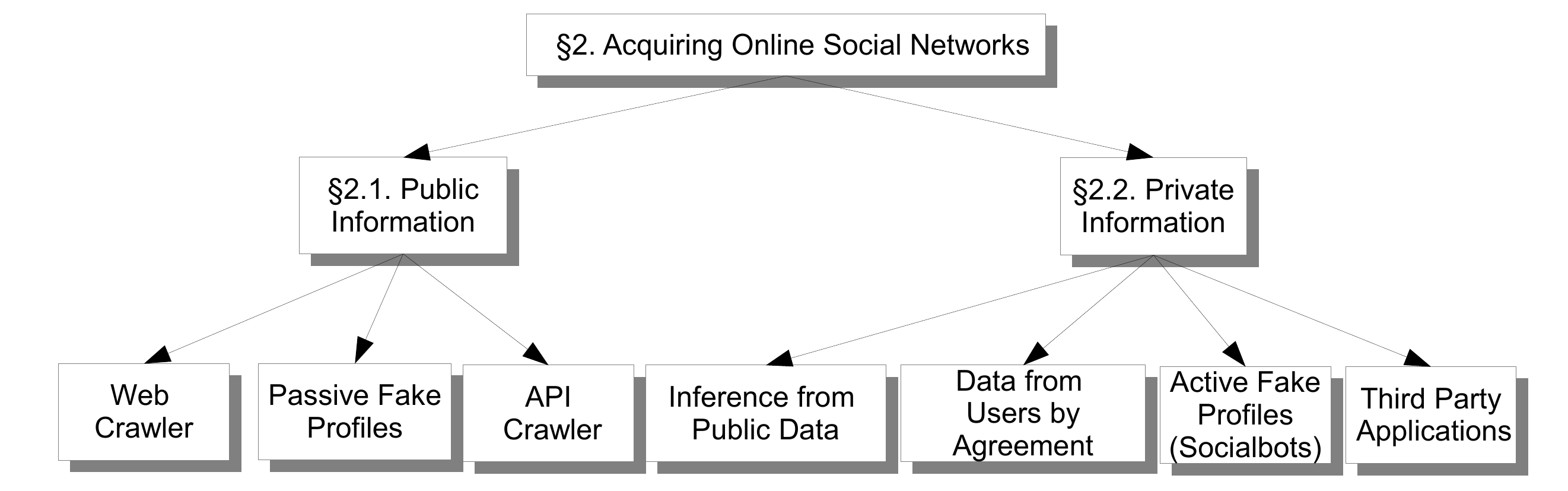}

	\end{center}	
\caption{Various OSN acquisition techniques.}
 \label{fig:crawling}
\end{figure*}

\s{With} the exponential growth of online social networks  usage \s{during the past several years}, many researchers have acquired OSN information for a range of purposes (See Table~\ref{tbl:osn_studies}), from studying the characteristics of large-scale online social network graphs (\cite{mislove2007measurement}) to improving road safety (\cite{fire2012data}). The OSN users' information can be divided into two main categories: public and private (see Figure~\ref{fig:crawling}). The public user's information \s{is data that} is available to all members of the OSN, while in many cases the user's private information is only accessible to friends of the user. For example, Facebook users' personal information is accessible to other users in the network according to each Facebook user's privacy settings (\cite{liu2011analyzing}). In some cases the user's information can be accessed only by \s{his or her}  Facebook friends, while in other cases the information can be accessed by every member in the network. In order to acquire OSN information, researchers \s{have} developed various techniques that aim to collect both OSN users' public and private information. In this section we present the different techniques by which researchers can acquire OSN information. In addition, we also present ethical issues which arise at the end of a study, after \s{completing} the acquiring process  \s{when,} the researchers want to share their acquired information with the academic community.
\subsection{Public Information}
To acquire OSNs users' public information, researchers \s{have primarily} 
used online web crawlers. These web crawlers can obtain the users' information by using the OSN's \s{application programming interface (API)} (\cite{mislove2007measurement,kwak2010twitter}) or by analyzing the raw data obtained directly from the OSN's web pages (\cite{fire2012data,fire2011link}). However, in some OSNs like Facebook, it is not possible to collect users' public information without being logged onto the OSN. To overcome this limitation, researchers have created passive fake profiles which are used to obtain access to the OSN public information (\cite{fire2012org,jernigan2009gaydar}). These fake profiles do not initiate friend requests to other users in the network and do not intervene in the OSN activity. By using \s{this method,} researchers are able to collect the OSN's public information with minimal intervening in the OSN activity and without accessing restricted users' private information. \s{The} results obtained from using these methods, \s{however, do} not give the researchers a wider picture of the OSN, which is crucial for certain type of studies, such as security and privacy studies. Moreover, some techniques not implemented in the right manner can abuse the OSN infrastructure and \s{negatively} affect the overall network performance. For example, using web crawlers, which initiate large amounts of rapid page requests, can create an overload on the OSN's servers, \s{resulting in  interference} with OSN activity.

\subsection{Private Information}
To acquire OSN users' private information, researchers have developed several techniques. These techniques include requesting private information directly from the users through applications and browser add-ons which integrate with the OSN (\cite{fire2012social,rahman2012efficient}); inferencing OSN users' private information by analyzing information obtained from \s{their} friends (\cite{jernigan2009gaydar, mislove2010you}); and even activating dynamic fake profiles, also known as socialbots, which initiate a series of friend requests in order to collect users' private information (\cite{boshmaf2011socialbot,elishar2012}). By using these methods, researchers can obtain a fuller picture of the studied OSN,  \s{including} private users' information. Moreover, these types of methods are extremely valuable when it comes to analyzing privacy and security issues in OSNs. For example, by using socialbots researchers were able to estimate the amount of private information exposed in Facebook to malicious fake users (\cite{boshmaf2011socialbot}). However, using these methods can also influence the OSN behavior and expose  personal sensitive \s{user} information, such as the user's sexual orientation (\cite{jernigan2009gaydar}). Moreover, fake profiles created by one research group may influence the research results of another research group that might treat them as real user profiles. Therefore, using these types of methods should be done with great care and with consideration to both the users' privacy and to the OSN's infrastructure.

\subsection{Sharing the Acquired Information}
After the networks' acquisition processes are finished, many researchers want to assist their peers by sharing their acquired datasets with the rest of the world. This is usually done by anonymizing the datasets and uploading them into dedicated websites (\cite{bgudata,konect,snap,Zafarani_Liu}). 
However, the sharing and anonymizing process \s{must} be done with great care and consideration to the OSNs users' privacy. Using the wrong anonymization methods can result in exposing the OSNs users' private and sometimes sensitive information (\cite{narayanan2011link,narayanan2006break, zimmer2010but}). 
Another ethical issue which needs to be taken into consideration is what to do with the acquired datasets after the study is over. Do the researchers need to delete and destroy the datasets\s{, or can they} store them in an encrypted manner for further use? Each action has its own positive and negative sides. If the researchers delete the datasets, they \s{fully} protect the OSNs users' privacy. However, they can not use them for further studies, and other researchers will not be able to compare their results to previous studies using the same data. If the researchers keep the \s{datasets, they} can jeopardize the OSNs users' privacy, but the datasets can be used for further studies without the need to \s{undertake the extensive} acquisition process again. We present our recommendations in Section~\ref{sc:sol} \s{for these issues}.



  

\begin{table*}[htb]
  \caption{\s{Various studies using different online social network acquiring methods.}}

  \centering

  \begin{tabular}{c}

      \includegraphics[width=0.85\textwidth]{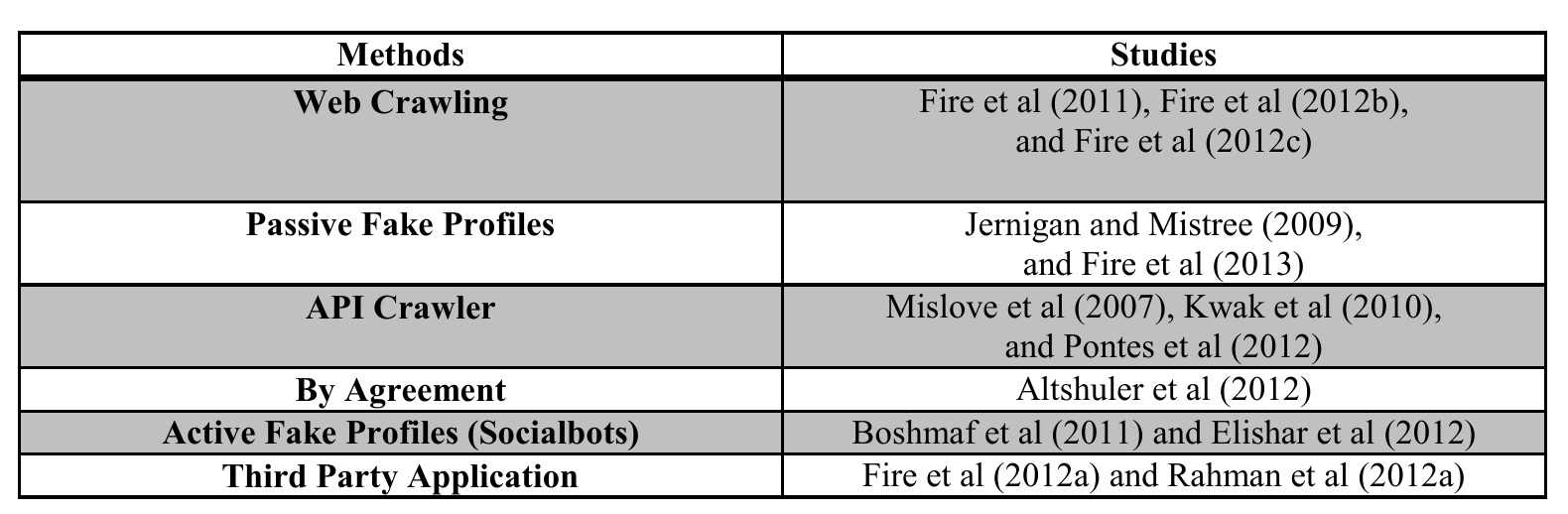}

  \end{tabular}

  \label{tbl:osn_studies}

\end{table*}

\section{Ethical Considerations} \label{sec:ethics} 
Online social networks offer a vast amount of data, useful for research in various disciplines for \s{both commercial and} academic purposes. In researching online social networks, a number of ethical challenges and dilemmas are \s{introduced} with respect to the involved entities. We consider the following as the  entities that may be impacted by \s{OSN research:} (1) the {\em users} of the OSN, (2) the {\em OSN operator}, and (3) the {\em advertiser/investor} in OSN. 

\begin{figure*}[th!]
	\begin{center}
			    \includegraphics[width=0.75\textwidth]{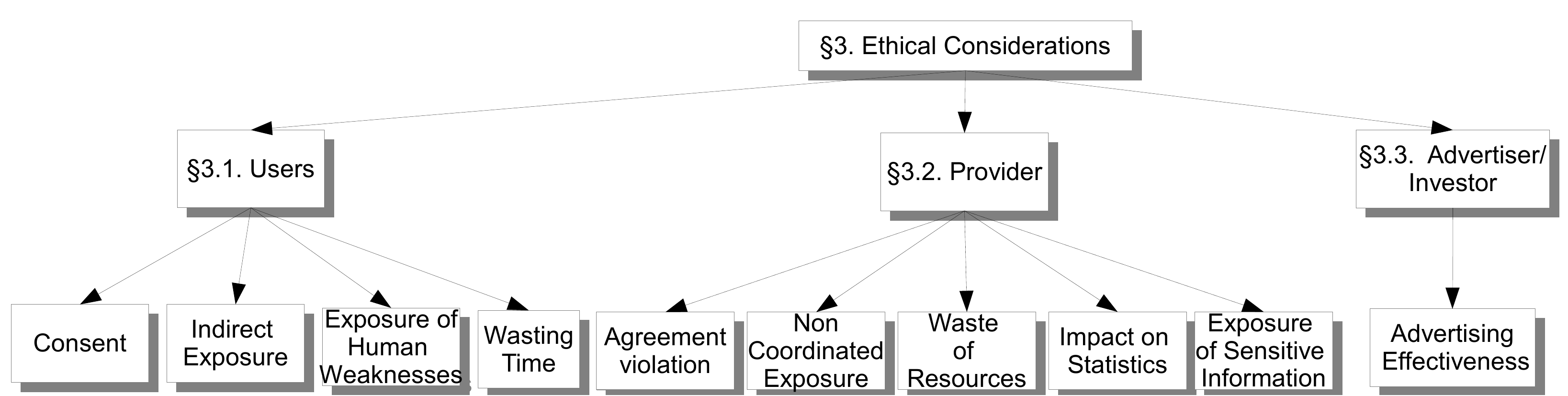}
	\end{center}	
\caption{Ethical considerations with respect to the user, the OSN, and the \s{advertiser/investor}  when performing research on an OSN.}
 \label{fig:ethics}
\end{figure*}

In this section we identify relevant ethical issues (see Figure~\ref{fig:ethics}). We then map the techniques in Section~\ref{sec:acquire} to the ethical problems that they pose (see Table~\ref{tbl:ethics}). 


\subsection{The Users}
Rightfully, ethical research considerations mostly focus on potential harm to the end users, and this also holds for OSNs. The main ethical issues concerning users \s{involve} their consent to participate in the experiment, and in particular, to \s{allow} any exposure of their private information. Within online social networks, users share lots of information with their \s{online connections}, and may make decisions based upon connections of their connections. This raises another ethical problem, \s{that of} indirect exposure, i.e., exposure of a user via a connection. 
\s{OSN research may also reveal} common user weaknesses, which could \s{then be exploited, and} they should be reported carefully to avoid being used against the users.  
\s{Finally, when users are unaware of an OSN experiment, yet affected by it,  there can be the ethical concern of potential loss of time. We next review each of these concerns.} 


\subsubsection*{Consent}
A basic ethical question is whether all users should be aware of being involved in an experiment, and if their consent \s{is required to participate in} the experiment and allow any resulting exposure of information. In many types of research involving humans, it is considered unethical to perform an experiment without consent.
\s{Indeed, many recent studies involve those who have explicitly volunteered, offering access to their personal online information 
(e.g. ~\cite{huber2011social})}.

On the other hand, many of the experiments on OSNs \s{have not required the} prior informing of users or the receiving of their consent. \s{In many such cases the research has} significantly contributed to society, such as identifying risks to OSN users and allowing the development of countermeasures, and the actual damage to individuals \s{seems negligible}. Can this justify performing research without a user's consent? 



\subsubsection*{\s{Indirect Exposure}}
A side effect of transitive trust is exposure of personal data to friends of a friend. 
\s{In addition, two friends may become connected who might not want to become connected.}

{\sc Exposing the Data of Friends of a Friend.}  \s{A user's consent}, such as for a ``friend request'' or to participate in research, does not \s{automatically} imply the consent of the user's friends. When performing research on an individual's data, the researcher also frequently gains access to personal information \s{about that individual's} friends.

{\sc Connecting $2^{nd}$ Level Friends.} Accepting a friendship request implies becoming visible to new users. Two users may not want to be connected via a third party; \s{they may wish} to hide the existence of their virtual profiles or want to keep the information that they post private. During \s{studies, researchers} may connect individuals profiles which would otherwise not be connected. For example, consider a bot (attacker controlled profile) that a researcher connects to\s{, not realizing that} it is a malicious profile whose goal is to constantly discover new users and to harvest information on these users. \s{This bot now has} a second degree access to all of the researcher's \s{OSN friends}, exposing their privacy. Furthermore, such a connection may also assist the bot in \s{finding even broader} connections, which it otherwise would not have the knowledge or tools to perform. For instance, often to \s{glean}  users' information, researchers resort to studies from sociology or physiology in order to construct their profile in such a way that increases the chances of other users accepting their friend requests; the attacker does not have the required knowledge to perform such a study on its own.

\subsubsection*{Exposure of \s{Human Weaknesses}} Research conclusions and results, if not disclosed carefully, can harm the user. Specifically, when conducting research that employs an OSN, the researcher may discover vulnerabilities pertaining to human weaknesses or to the user-OSN interaction that were not known before. For instance, if a researcher discovers that a friendship request from a profile with a photograph of an attractive female is more likely to be accepted by OSN users, then he \s{should be extremely careful to} not disclose this prior to notifying the OSN provider \s{(see \cite{boshmaf2011socialbot})}. As a countermeasure, the OSN provider could, for instance, display a warning message to a user when detecting certain profile properties that are known to be suspicious.


\subsubsection*{Wasting Time}
The process of luring a user into accepting friend requests may be time-consuming  not only for the researcher but also for the user.

\subsection{The OSN Operator} Using the platform that an OSN provides for research may stand in violation of the OSN's terms of service. \s{This}  may further result in potential harm to the OSN or to its users, e.g., due to wasted resources, exposure to competition, or a reduced customer base.  \s{Several ethical considerations relating to OSN operators are described below.}

\subsubsection*{Agreement Violation}
Creating fake accounts often stands as a violation of conditions of an OSN, which forbids creating more than one account or using fake accounts. 
Fake accounts may expose the OSN to lawsuits or harm its reputation.

\subsubsection*{Non-Coordinated Exposure}
 OSN platforms may have vulnerabilities and can even be used as a vector in launching attacks, thus harming users, \s{hurting} other networks and services on the Internet, and exhausting the resources of the OSN. For instance,~\cite{huber2011friend} discovered a new attack: the friend-in-the-middle attack that can be used to harvest social data in an automated fashion. When conducting a study on \s{online social networks,} researchers often discover such vulnerabilities. It is important\s{, therefore,} to establish a procedure whereby the researchers can publish the OSN vulnerabilities and attacks and take the necessary precautions, e.g., contacting the OSN and allowing it to patch the exploit to prevent abuse by malicious parties. If a vulnerability is exposed without coordination with the OSN, it may be \s{subsequently}  exploited by attackers to \s{compromise} the OSN and its users. 
 
As an example, recently \cite{athanasopoulos2008antisocial} showed that malicious users can take control of the social network visitors  by remotely manipulating their browsers through legitimate web control functionality, e.g., using image-loading HTML tags. They also demonstrated that Facebook users can be exploited as a vector in launching a denial of service attacks. Clickjack attacks can hijack users' web sessions (\cite{rydstedt2010busting}) if the OSN does not employ sufficient countermeasures.

It is important to emphasize that research on OSNs is \s{clearly valuable} and should be encouraged since weaknesses \s{as described  above} may be exploited by malicious attackers without the awareness of the OSN, the users, or the research community.  Furthermore, research allows \s{the development} of patches and countermeasures to preventing the vulnerabilities from being exploited.

\subsubsection*{Waste of Resources}
Creating fake profiles consumes resources on the OSN, including storage, communication, and processing.

\subsubsection*{Impact on Statistics}
Fake accounts bias statistics and may provide misleading information on trends, \s{resulting in wide-ranging}  commercial implications.

\subsubsection*{Exposure of Sensitive Information}
\s{While running a} study, researchers may discover sensitive information pertaining to the OSN, such as the way its algorithms work. Exposure of this information may benefit the OSN's competitors and \s{consequently}  harm the OSN and produce a negative commercial impact.

\subsection{The Advertiser/Investor}
Fake accounts, created during \s{a research study}, may influence the perceived popularity of an \s{online social network; that is, the experiment may increase the OSN's share value through} impressing the shareholders, or \s{it may influence} the effectiveness of advertising on the OSN, which may not translate into a profitable and sustainable business. According to the analytics service of Limited Run, an online shopping platform provider, the ad system on Facebook is not reliable, and $80\%$ of the ad clicks come from fake accounts or bots, which drive up the advertising costs (\cite{limitedrun}). Furthermore, if the investor (or the advertiser) pays per profile, fake accounts \s{artificially} inflate the value of the OSN. For instance, in August 2012 Facebook was reported to have more than 83 million fake profiles, which \s{are about} $8.7\%$ of the total number of profiles (\cite{facebook_fake}); this indicates a \s{notable} growth of fake accounts from $6.0\%$ in March 2012. 


\begin{table*}[t]

    \caption{Mapping between the ethical considerations and the research methods deployed for studies on OSNs.}

  \centering

  \begin{tabular}{c}

      \includegraphics[width=0.70\textwidth]{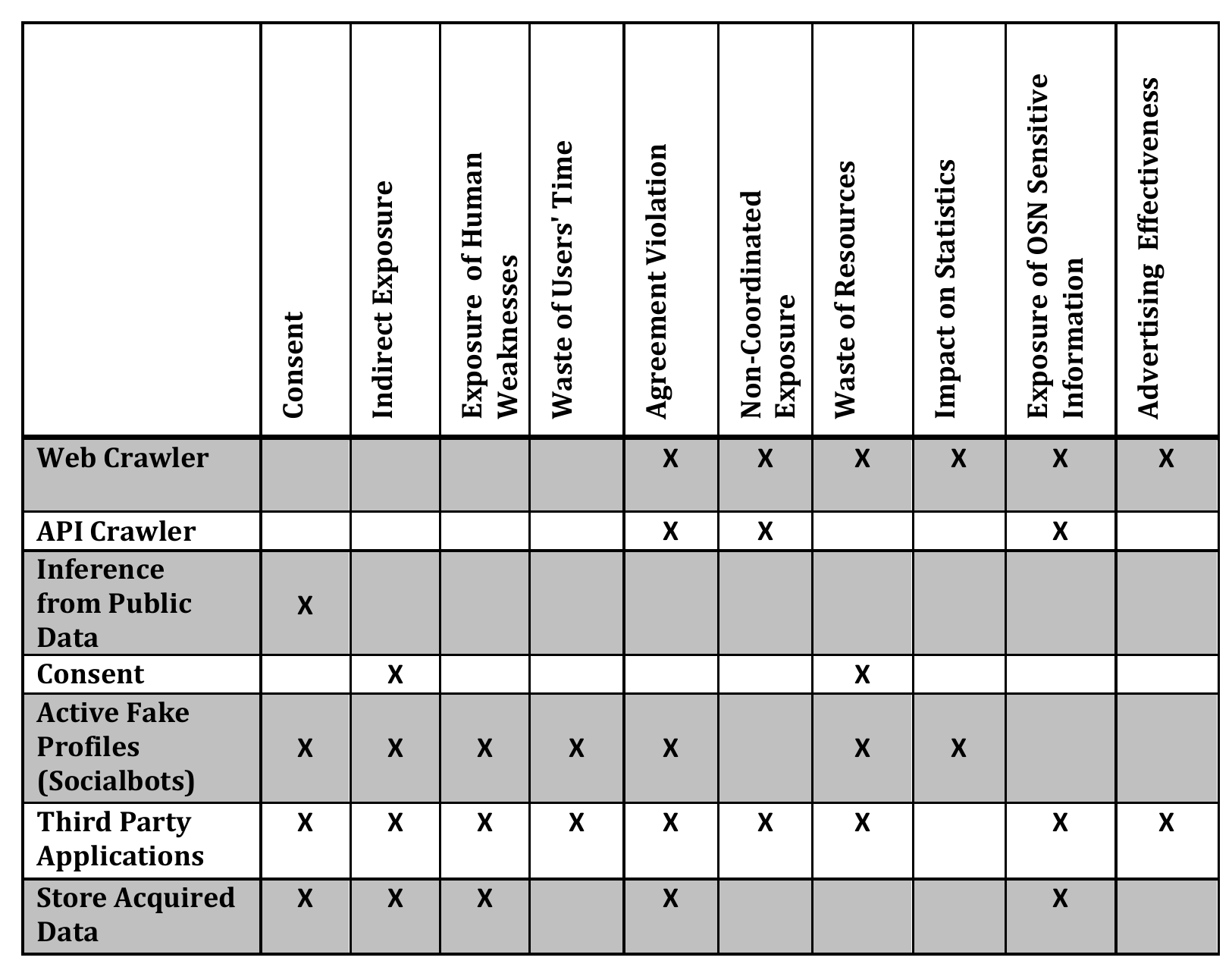}

  \end{tabular}

	    \label{tbl:ethics}

\end{table*}

\section{\s{Ethical Research Recommendations}}\label{sc:sol}
In the previous section, we discussed a large set of ethical issues with OSN research methodologies.
In this section, we explore different solutions, balancing the ethical principles of {\em confidentiality}, {\em anonymity}, {\em informed consent}, and {\em avoidance of disruption and waste}, \s{with} the benefits from experimental research on deployed OSNs. 

The ethical concerns presented in Section~\ref{sec:ethics} can be divided among {\em concerns related to the harvesting} of data from the OSN; {\em concerns related to the exposure of data}; {\em concerns about  the impact on the structure, availability, and security} of the OSN; and finally {\em concerns about the conduct of researchers}. We discuss each of these sets in the following subsections; before that, in the next subsection, we discuss the specific issue of consent. 


\subsection{Consent}
We begin with some general recommendations, observations, and proposals. 

We first discuss the {\em informed consent} concern and possible remedies. 
Obtaining user consent in OSN research has two main challenges: (1) researchers may be unable to know the real age of the user in order to verify that the user is of age (for informed consent), and (2) \s{the very awareness of having informed consent may influence a user's behavior, impacting the subsequent research on user behavior itself}. 

We do not know of a truly reliable solution to the age problem since many OSNs only allow  adult users, yet many under age users register anyway, \s{simply} providing an incorrect age. We believe this issue is not specific to research, as this problem relates to the actual provision of OSN services\s{; therefore it is } reasonable to ignore this concern and for researchers to use the stated age of the user (usually available). 

The user-awareness problem is much more severe for research related to user behavior. We suggest two solutions: 
\subsubsection*{Long-Term Research.} A user's awareness of \s{having been} measured in some experiment is quickly eroded after \s{using a system for some ``real'' purpose. Hence, the effect of user awareness can be minimal in research that is long term}. This was confirmed by \cite{conf/esorics/HerzbergM11} and used in experiments in secure usability. 

\subsubsection*{\s{Post-Research Informing/Compensating.}} Another \s{option} is to inform users of their involvement in research {\em after} the measurements \s{have been completed, even though} the users did not express their (informed) consent. It may be appropriate to also offer some token of compensation as well as an explanation of the need to use this method (instead of informed consent) and \s{to emphasize} the social value of the research.

This is clearly not as ethical as proper informed consent. However, this method may still have some advantages, in particular: (1) users have the ability to provide feedback, providing some indication of the amount of harm, \s{if any, that they have experienced}, and (2) this provides a {\em negative training function}, helping users to become more savvy  at detecting OSN fraud; see  \cite{conf/esorics/HerzbergM11}.


\subsection{Harvesting Process}

As described in \s{Section~\ref{sec:ethics}}, the harvesting process \s{can affect} OSN performance, \s{influence an OSN's} reputation, and threaten users' privacy. The harvesting is performed directly by crawlers or indirectly by fake identities. One of the first ways to reduce the ethical concerns \s{related to} harvesting is to improve the sharing of datasets among the researchers such that not every researcher will harvest his own data. Currently, the information retrieval community has created many datasets that are being shared by all the researchers in the community (\cite{letor}). The social network research community should adopt this strategy in order to reduce harvesting as much as possible. Alternatively, perhaps the social network operators should create datasets \s{specifically} for researchers (``if you cannot fight them, join them'') after informing their users about the new policy. The social network operators are in a better position to apply advanced anonymization techniques than any small research group.

\subsection{Research Results}
Research results may influence both the users and operators of OSNs as described \s{previously}.
\s{As with many types of research, vulnerabilities} are routinely discovered and often disclosed before the operators have had the chance to patch up the
problem. This \s{issue is} exacerbated with respect to OSN platforms and their users since each weakness
has the potential to affect an enormous number of parties,
with detrimental results.
Due to the importance \s{of the potential data exposure,} we recommend setting up a Coordinated Emergency Response Team (CERT) for vulnerability
disclosures. Such an entity will allow all of the involved \s{parties to be quickly notified}, and it will give them an opportunity
to patch the problem before the exploit becomes public. Researchers should also suggests countermeasures to the attack
and inform the \s{operators accordingly}.

Experiments that involve human subjects should require a lengthy
approval process, whereby it is verified that the research
can cause no harm. Since OSN research involves human subjects, or more specifically their data, an approval
process should be required for cases in which new research methods are being implemented.

A potential reaction of the OSN operator, e.g., due to concern for legal actions against it, may be to prevent any kind of research on the OSN. However, we caution against this reaction, noting that such a {\em security by obscurity} is risky and \s{can endanger} both the users of the OSN and the OSN platform (as well as other potential third parties). Vulnerabilities may be discovered and abused by malicious attackers, instead of \s{used beneficially by} legitimate researchers. \s{Attackers} do not publish the vulnerabilities and attacks; researchers, on the other hand, create awareness \s{and provide solutions to these} vulnerabilities. 
\s{Therefore, we recommend that the OSN operators address these issues and recognize the legitimate need for research to provide appropriate, beneficial contributions.}

\subsection{Influence on the Network}
\s{In general, OSN operators} need to accept the fact that researchers will \s{undoubtedly continue to} create fake profiles, and the \s{operators'} algorithms will
never be able to detect all of them. Therefore, the OSN operator should allow fake profiles \s{to be created}, but it should also \s{offer a specified} procedure whereby a researcher can notify the OSN and request approval (possibly anonymously) for \s{introducing} profiles that are fake and dedicated to a specific study, indicating relevant information such as the duration of the study and purpose of the research. The OSN should have an approval (or alternately, rejection) procedure for each profile, along with an appropriate \s{method to notify}  the researcher. The OSN may also request an approval or certificate from the researcher, stating that the study is ``ethical.''
It is important to note that the proposed procedure would allow researchers to spend fewer resources on having to generate fake profiles that cannot be detected by the OSN operator.

We recommend that generally the OSN should allow the requests since good research will help protect the OSN platform and its users. 
However, it may be reasonable to limit the number of fake profiles to some \s{pre-determined threshold at any one time}.


\subsection{Researchers' Conduct}
We recommend that researchers use the online social networks' APIs when possible, and then resort to fake {\em passive} profiles (instead of active) \s{if needed}. The identifiers of the profile  \s{(e.g., a photo)} should not expose the identity of the \s{individual}. When the study concludes,  the researcher should remove the fake profile and, \s{if possible}, inform users connected to that profile that it was a fake profile for research purposes. \s{When possible, the researcher should also provide} information and token compensation, and allow users to provide feedback. This has multiple benefits: (1) fairness to users; (2) better awareness of the possible harm to users, allowing improved ethical decisions in future research; (3) reduction in harm to users; (4) \s{potentially highly significant negative training} to users, teaching them to be more cautious and to be able to identify fake identities in the future (see \cite{conf/esorics/HerzbergM11}). 
 
Researchers should use cryptographic and other techniques for anonymization of the data as well as for confidentiality. The researchers should handle the study data, as well as the findings, with great care. Following the research, the data should be removed, or if it is required for future research, it should be stored encrypted. When sharing the data, we recommend applying appropriate anonymization techniques \s{which remove} all identifying data that can leak information when combined with other sources. 
\s{See (\cite{zimmer2010but}) for further information on such deanonymisation attacks}.
When transferring the data over the network and when accessing the OSN, encryption should be used. 
When conducting the study, the researchers should also consider the load on the OSN and the overhead to users. 
Finally, the researchers should maintain precise and accurate records of the data collected, the profiles created, the people involved in research, and \s{any other pertinent information. This must be preserved in a secure manner}. 

\section{\s{Previous Research and Ethical Considerations}}
\label{sc:prev}

In this section, we will use previous studies that employed fake identities \s{to illustrate} that it is possible to achieve similar research results, but with lower impact to the OSN members' privacy and to the OSN operator resources.

\cite{fire2012org} presented a method for mining \s{an organization's} topology through the use of passive fake identities which collected employees' public information from their Facebook profiles. 
\cite{elishar2012}  employed active fake identities to achieve the same goal of mining organizational topology. 
\s{\cite{elishar2012} initiated} friend requests to Facebook users who worked in a targeted organization. 
Upon accepting these friend requests, users unknowingly exposed information about themselves and about their workplace.
\s{This technique} was tested on two real organizations and successfully infiltrated both. 
Compared to the \s{passive fake identity method, the technique utilizing} active fake identities, was able to discover up to 13.55\% more employees and up to 18.29\% more informal organizational links. 
However, similar results were achieved from both studies when identifying leadership roles using different centrality measures. 
Hence, it is \s{possible to infer} leadership roles without the need to employ active fake profiles, which can compromise the OSN users' privacy and impact the OSN operator resources.

\cite{fire2012strangers} presented a novel method for the detection of fake identities in OSNs by \s{using only} the social network's own topological features. Reliance on these features alone ensures that the proposed method is generic enough to be applied on a range of social networks. In order to train a \s{machine learning} classier, training had to be created and collected. The training set had to include labeled fake identities. One way to create a training set is to add fake identities to an OSN and then collect the modified network as a labeled training set. To avoid modifying the OSN \s{itself}, which will create a negative impact on the OSN operator, a different approach was taken. To create positive examples for the classifiers, Fire et al developed a code which simulated the infiltration of a single fake identity (or a group of fake identities) to social networks. For each social network, the simulator loaded the topology graph and inserted 100 new nodes, which represented 100 fake identities. The insertion process of each fake identity into the graph was done by simulating a series of ``follow'' requests sent to random users in the network. Each fake user had a limit on the number of friend requests in order to comply with a reality in which many social networks limit the number of user requests allowed for new members (exactly for the purpose of blocking spammers and socialbots).

\section{Conclusions}
\label{sc:conclusions}
%

Research \s{involving} the operation of  widely-used \s{online social networks the behavior} of  OSN users, is vital to the design of improved OSNs, OSN applications, and especially OSN privacy and security mechanisms. 
\s{Indeed, research on online social networks and OSN data can be useful in various research disciplines, such as the study of social behavior and the design of secure and usable social networks. It can also have value and applications in real life and; for example, it can facilitate detection and a priori prevention of terror attacks, theft, coercion, and other critical situation.}

\s{Many commonly used} research techniques\s{, however,}  raise ethical concerns, which are rarely even considered by researchers or the community at large. 
\s{For instance, such research can jeopardize users' privacy as well as expose potential vulnerabilities of the OSN platform to abuse by malicious attackers.
A trivial solution is to limit such research to use on simulated environments. However, in order for research on OSNs to be of practical relevance and \s{maximum usefulness} it must be conducted \s{under} realistic conditions, \s{using} real users' data. In particular, in contrast to other research topics, \s{studying} a simulated OSN may not yield realistic results, especially \s{if the entire research goal} is to analyze and test real users' data and interactions. 
}

With this paper, we strive to draw attention to this important issue and initiate a discussion within the OSN research community \s{to encourage the adoption} of an appropriate code of ethics, \s{which can then be utilized} in agreements and technical \s{standards. This code would be a means to} facilitate good research, with acceptable trade-offs between ethical considerations and the social benefits of precise, available, and timely research on these important issues.   
\s{As we point out in this work, an effort from both the research community and industry is required to define and standardize mechanisms that will enable this legitimate and significant research while ensuring privacy of OSN users and their data. A number of such mechanisms that can facilitate ethical research on OSNs have been outlined. As online social networks continue to gain prominence, ethical considerations become increasingly critical.
}

\bibliographystyle{IEEEtran.bst}
\bibliography{osm}

\end{document}